\documentclass[
    aps,
    prd,
    nofootinbib,
    superscriptaddress,
    10pt,
    notitlepage,
    preprintnumbers,
    twocolumn]{revtex4-2}

\usepackage{anyfontsize}
\usepackage{amssymb}
\usepackage{amsmath}
\usepackage{amsfonts}
\usepackage{amsbsy}
\usepackage{multirow}
\usepackage{newtxtext}
\usepackage[varvw,smallerops]{newtxmath}

\DeclareSymbolFont{CMlargesymbols}{OMX}{cmex}{m}{n}
\DeclareMathDelimiter{(}{\mathopen} {operators}{"28}{CMlargesymbols}{"00}
\DeclareMathDelimiter{)}{\mathclose}{operators}{"29}{CMlargesymbols}{"01}
\DeclareMathDelimiter{[}{\mathopen}{operators}{"5B}{CMlargesymbols}{"02}
\DeclareMathDelimiter{]}{\mathopen}{operators}{"5D}{CMlargesymbols}{"03}

\usepackage{siunitx}
\usepackage{booktabs}
\usepackage{tensor}
\usepackage{mathrsfs}
\usepackage{bm}
\usepackage{microtype}
\usepackage[utf8]{inputenc}

\usepackage{graphicx}
\usepackage{epsfig}
\usepackage{epstopdf}

\usepackage[linktocpage,breaklinks]{hyperref}
\usepackage[dvipsnames]{xcolor}

\definecolor{romared}{RGB}{142,0,28}
\hypersetup{colorlinks=true,
            citecolor=romared,
            linkcolor=romared,
            urlcolor=romared}

\newcommand{\be}{\begin{equation}}
\newcommand{\ee}{\end{equation}}

\newcommand{\del}{\partial}

\def\be{\begin{equation}}
\def\ee{\end{equation}}
\newcommand{\beq}{\begin{eqnarray}}
\newcommand{\eeq}{\end{eqnarray}}

\begin{document}

\title{Black hole quasinormal mode excitation and mixing for coupled fields}
\author{Caio F. B. Macedo} \email{caiomacedo@ufpa.br}
\affiliation{Programa de Pós-Graduação em Física, Universidade Federal do Pará, 66075-110, Belém, PA, Brazil}
\author{Orival R. de Medeiros}
\email{orival@ufpa.br}
\affiliation{Programa de Pós-Graduação em Física, Universidade Federal do Pará, 66075-110, Belém, PA, Brazil}
\author{Emanuele Berti}
\affiliation{William H. Miller III Department of Physics and Astronomy, Johns Hopkins University,
3400 North Charles Street, Baltimore, Maryland, 21218, USA}
\author{Hector O. Silva}
\affiliation{Department of Physics and Illinois Center for Advanced Studies of the Universe,
The Grainger College of Engineering, University of Illinois Urbana-Champaign, Urbana, Illinois 61801, USA}
\author{Sebastian H. V\"olkel}
\affiliation{Theoretical Astrophysics, IAAT, University of T\"ubingen, Auf der Morgenstelle 10, D-72076 T\"ubingen, Germany}

\begin{abstract}
The idea of testing the nature of black holes and possible deviations from general relativity from their oscillation spectrum, known as ``black hole spectroscopy'', has become an experimental science.
In many modified theories of gravity, the gravitational perturbations couple to additional  degrees of freedom, leading to mode mixing and to a more complex ringdown signal.
We adopt the parametrized quasinormal mode formalism for nonrotating black holes to study coupled perturbations in modified gravity theories in order to understand how the interaction with additional fields affects the quasinormal mode frequencies and their amplitudes. We compute the quasinormal mode spectrum of the coupled system in the frequency domain using direct integration and a matrix-valued continued-fraction method. We then compare the results with time-domain evolutions, where we extract the mode frequencies and amplitudes using the Prony method as well as least-squares fitting. In the weak-coupling regime, the scaling of the amplitudes associated with the gravitational-led and secondary-led modes can be understood perturbatively. The excitation of additional families of ringdown modes can become significant even when the frequency shifts remain moderate. We use our numerical time evolutions to provide ready-to-use coefficients for the mode amplitudes for representative initial data configurations. We work out specific examples, including dynamical Chern-Simons gravity and the special near-resonant case of coupled perturbations between the polar and axial quasinormal modes. Our theory-agnostic characterization of coupled systems should be useful to quantify systematic effects in black hole spectroscopy with next-generation gravitational-wave detectors.
\end{abstract}

\date{\today}
\maketitle

\section{Introduction}\label{sec:intro}

The detection of gravitational waves has opened a new era in astrophysics and cosmology, providing a novel observational window to probe the Universe via spacetime dynamics, alongside conventional electromagnetic methods~\cite{Berti:2015itd,Berti:2018cxi,Berti:2018vdi}. The first direct observation of gravitational waves from GW150914~\cite{LIGOScientific:2016aoc,LIGOScientific:2016vlm,LIGOScientific:2016lio} marked a  milestone that ushered in a new era of precision tests of the dynamics of the Kerr geometry~\cite{Kerr:1963ud,Vishveshwara:1970zz,Press:1971wr,Detweiler:1980gk,Leaver:1986gd,Dreyer:2003bv,Berti:2005ys,Berti:2007zu,Brito:2018rfr}. This first event had a signal-to-noise ratio (SNR) $\approx 7$ in the so-called ``ringdown'' relaxation phase dominated by a superposition of quasinormal modes (QNMs), and therefore black hole spectroscopy tests were inconclusive: see e.g., Refs.~\cite{Isi:2019aib,Cotesta:2022pci,Finch:2022ynt,Isi:2023nif,Carullo:2023gtf,Correia:2023bfn,Wang:2023ljx}, and Ref.~\cite{Berti:2025hly} for a review.
The ``twin'' event GW250114, observed about ten years later, has a ringdown SNR comparable to the total SNR of GW150914~\cite{LIGOScientific:2025rid} and allowed for the first spectroscopic test of the Kerr nature of the remnant~\cite{LIGOScientific:2025wao,LIGOScientific:2025rid}.
Next-generation detectors, such as the Laser Interferometer Space Antenna (LISA)~\cite{LISA:2017pwj}, the Einstein Telescope~\cite{Punturo:2010zz,ET:2025xjr} and Cosmic Explorer~\cite{Reitze:2019iox,Evans:2023euw}, will provide even higher SNR measurements and enable precision black-hole spectroscopy~\cite{Berti:2016lat,Bhagwat:2019dtm,Maselli:2019mjd,Barausse:2020rsu,Evans:2023euw,ET:2025xjr,Maselli:2023khq,Yi:2024elj}.
To fully exploit these capabilities, it is essential to understand not only the instrumental sensitivity, but also the detailed morphology of the signal itself.

If the assumptions underlying the so-called ``no-hair theorem'' and GR hold, independently measured QNMs must yield consistent values of the black hole's mass and spin~\cite{Detweiler:1980gk,Leaver:1986gd,Dreyer:2003bv,Berti:2005ys,Berti:2007zu,Brito:2018rfr}. 
Deviations from this consistency test would indicate either modifications to the background spacetime geometry or dynamical corrections arising in theories beyond general relativity (GR).

A common aspect to many viable modified theories of gravity is the presence of additional fields. In black hole perturbation theory, this generally yields coupled perturbation equations with additional QNM spectra and mode couplings.
Standard ringdown tests usually focus on extracting gravitational QNMs, and do not consistently include the dynamics of coupled perturbations.
While some studies have begun to address the interesting dynamics of coupled systems in specific modified gravity theories (see e.g.~\cite{Cardoso:2009pk,Molina:2010fb,Blazquez-Salcedo:2016enn,Glampedakis:2019dqh,Silva:2019scu,Bryant:2021xdh,Silva:2024ffz,Silva:2026jih,Blazquez-Salcedo:2024oek,Khoo:2024agm,Blazquez-Salcedo:2024dur,Chung:2024ira,Chung:2024vaf,Chung:2025gyg,Chung:2025ucb,Chung:2025ucb,Cano:2024wzo,Cano:2025mht,Cano:2025ejw,Bah:2026aia,Weller:2026jqu,
Crescimbeni:2024sam,Motohashi:2024fwt,Cavalcante:2024swt,Cavalcante:2024kmy,Yang:2025dbn,Takahashi:2025uwo,PanossoMacedo:2025xnf,Cavalcante:2025abr,Hu:2025efp,Lestingi:2025jyb,Asuncion:2026efk,Imafuku:2026rpn} for a nonexhaustive list)
and in parametrized frameworks~\cite{Glampedakis:2017dvb,Cardoso:2019mqo,McManus:2019ulj,Kimura:2020mrh,Volkel:2022aca,Cardoso:2024mrw}, a systematic, theory-agnostic characterization of the impact of mode mixing on ringdown observables is in its infancy: for example, Ref.~\cite{Crescimbeni:2024sam} placed bounds on scalar QNMs in gravitational-wave data.

In this work, we study the implications of the coupling of multiple fields in the ringdown, both in theory-agnostic and theory-specific examples. Due to the significant technical challenges associated with perturbations of rotating black holes, we focus on nonrotating backgrounds for simplicity. Our goal is to derive the main qualitative features and some semianalytic results within the parametrized QNM framework.

The main idea of this parametrization is to introduce phenomenological deformations to the effective potentials governing black hole perturbations, schematically written as $V = V_{\rm GR} + \delta V$, where the correction term is expanded in powers of $1/r$ and controlled by a set of coefficients that encode deviations from GR.
The parametrized QNM framework, initially developed in the frequency domain~\cite{Cardoso:2019mqo,McManus:2019ulj,Kimura:2020mrh,Volkel:2022aca,Cardoso:2024mrw}, was recently extended to the time domain~\cite{Thomopoulos:2025nuf,deMedeiros:2025ayq}.
Lower-order deformations primarily modify the late-time Price tail and can generate quasibound states, whereas higher-order terms directly affect the ringdown waveform itself. The perturbative stability of the fundamental mode, together with the agreement between analytical predictions and numerical evolutions, show that the parametrized QNM framework provides a robust model-independent approach to study deviations from GR.

The parametrized framework for coupled fields involves parametrized coupling functions and enables theory-agnostic studies of QNMs when gravitational perturbations interact with (presumably subdominant) extra fields~\cite{McManus:2019ulj}. Even in the simplest case of two fields, the coupling induces \textit{mode-mixing}: an initial perturbation in one field excites both of them, transferring energy between the different classes of perturbations and contaminating the purely gravitational ringdown signal with the characteristic frequencies of the secondary field.

We show that, in the weak-coupling limit, the system can be solved perturbatively order-by-order in the coupling, although the mere presence of multiple modes in both sectors is already sufficient to alter the expected ringdown morphology.
We investigate and quantify this ringdown signal contamination in coupled systems by applying the parametrized QNM framework of Refs.~\cite{Cardoso:2019mqo,McManus:2019ulj} to study the scaling of the mode amplitudes in the time domain. By analyzing mode-mixing and comparing different numerical techniques, such as the continued-fraction (CF) method in the frequency domain and time-domain fitting techniques such as the Prony method, we characterize the effects of the coupling on the QNM spectrum.
We also find that the presence of coupling functions related to near-horizon modifications affects the late-time power-law tails.
The main message of this work is that the coupling induces interesting features that go beyond a simple shift of the gravitational QNM frequencies, and that must be better understood to lay the groundwork for next-generation black hole spectroscopy.

This paper is organized as follows. In Sec.~\ref{sec:framework}, we present the parametrized framework for coupled perturbations, including the master equations, coupling structure, and numerical methods used in both frequency- and time-domain calculations. In Sec.~\ref{sec:results}, we present the QNM spectra and time-domain evolutions, focusing on mode excitation, mixing, and coupling-dependent amplitudes, and we discuss illustrative examples, including dynamical Chern-Simons (dCS) gravity and the degenerate Zerilli/Regge-Wheeler system.
Finally, in Sec.~\ref{sec:conclusion}, we summarize our results and discuss future directions. Additional details on our frequency-domain calculations and on the Prony method are provided in Appendices~\ref{app:frequencies} and~\ref{app:Benchmarking}, respectively. Throughout this paper we use geometrical units ($G=c=1$).

\section{Parametrized framework: Coupled fields}\label{sec:framework}

We consider a parametrized framework describing the coupling between gravitational perturbations and extra fields propagating on a Schwarzschild background; this assumption does not necessarily limit us to theories that admit only the Schwarzschild solution, see e.g., Ref.~\cite{Franchini:2022axs}.
The dynamics of the field is governed by a system of coupled wave equations for the variables $\Psi_s$, where the index $s$ labels the spin of the field.
We will focus on the coupling between fields of integer spin $s=0,\,1,\,2$,
corresponding to scalar, vector, and gravitational perturbations, respectively.

The perturbation equations in matrix form read~\cite{McManus:2019ulj}
\begin{equation}
    \left[\frac{\del^2 }{\del x^2}- \frac{\del^2 }{\del t^2}-\mathbf{V}(r)\right]\mathbf{\Psi}(t,x)=0\,,~\label{eq:matrix}
\end{equation}
where $\mathbf{\Psi}$ is an $N$-dimensional column vector representing a collection of the master variables $\Psi_s$, and $\mathbf{V}$ is an $N\times N$ matrix that generalizes the effective potential by including off-diagonal interaction terms between the perturbations.
Following Ref.~\cite{McManus:2019ulj}, we define the tortoise coordinate, $x$, through ${\rm d}x={\rm d} r/(1-2M/r)$.
All leading-order modifications to GR are encoded in the effective potential matrix $\mathbf{V}$.
In the more general case where we consider multiple fields of different spins, or different parity sectors of perturbations with same spin that couple to each other, an additional label may be necessary to identify the master variables.

For simplicity, we focus on the case of two coupled fields; the extension to three or more fields will be presented elsewhere.
We are particularly interested in the case in which one field is large and can therefore be treated as ``primary,'' while the ``secondary'' field remains small.
In black hole spectroscopy, we typically expect gravitational perturbations to correspond to the primary field. The small secondary field would correspond to additional degrees of freedom in modified theories of gravity.
The potential matrix then reduces to:
\begin{equation}
    \mathbf{V}=\left(
    \begin{array}{cc}
        V_{s_1}(r) & \mathcal{C}(r) \\
        \mathcal{C}(r) & V_{s_2}(r)
    \end{array}\right),
\end{equation}
where we wrote $V_{12}(r)=V_{21}(r)=\mathcal{C}(r)$, so that the potential matrix is symmetric. We adopt this form for the coupled-field system considered here. 
The quantities $s_1$ and $s_2$ denote the spins of the perturbations associated with the primary and secondary fields, respectively.
Unless stated otherwise, from now on we will mainly consider the case in which $s_1=2$ and $s_2$ is either a scalar ($s_2=0$) or a vector ($s_2=1$) field.

The effective potential for decoupled, massless perturbations of spin $s$ is
\begin{equation}
	V_s(r) = f(r)\left[
	\frac{\ell(\ell+1)}{r^2} + (1 - s^2)\frac{2M}{r^3}
	\right],
	\label{eq:Vef}
\end{equation}
where $f(r) = 1 - 2M/r$.
Equation~(\ref{eq:Vef}) reduces to the Regge-Wheeler potential when $s=2$, and to the scalar and vector potentials when $s=0$ and $s=1$; see e.g., Refs.~\cite{Regge:1957td,Zerilli:1970se,Zerilli:1970wzz,Berti:2009kk}.
The evolution equations are given by
\begin{subequations} \label{eq:Psis1s2}
\begin{align}
	\left[ \frac{\del^2 }{\del x^2}- \frac{\del^2 }{\del t^2} - V_{s_1}(r) \right] \Psi_{s_1}
	&= \mathcal{C}(r)\, \Psi_{s_2}\,,\label{eq:Psis1} \\
	\left[ \frac{\del^2 }{\del x^2}- \frac{\del^2 }{\del t^2} - V_{s_2}(r) \right] \Psi_{s_2}
	&= \mathcal{C}(r)\, \Psi_{s_1}\,.\label{eq:Psis2}
\end{align}
\end{subequations}

We parametrize the coupling between the fields through the function~\cite{McManus:2019ulj}
\begin{equation}
	\mathcal{C}(r) =\sum_{j=3}^N f(r)\,\frac{\alpha_j}{4M^2} \left(\frac{2M}{r}\right)^j,
    \label{eq:mixing}
\end{equation}
where the dimensionless coefficients $\alpha_j$ control the strength of the interaction. Notice that we allow for different coupling constants $\alpha_j$ at each order in the expansion. We consider modifications starting with $j=3$, such that the late-time tail behavior of the perturbations is not affected by the coupling.
We also focus on specific values of $j$~\cite{Cardoso:2019mqo,McManus:2019ulj,deMedeiros:2025ayq}. This allows us to extract generic trends for each power of $1/r^j$, which can be thought of as the leading-order modification of different coupling scenarios in which the interaction gets stronger near the black hole horizon.

The structure of the coupled system is inspired by previous studies of modified gravity theories. For example, similar coupled wave equations arise in dCS gravity, where axial gravitational perturbations couple to a scalar field through a parity-violating term~\cite{Molina:2010fb}.
The parametrized framework here captures the essential features of such couplings in a model-independent way, and further allows us to consider coupling to scalar and vector fields.

In our numerical analysis, we will mostly focus on the axial (Regge-Wheeler) function, although we will also make some remarks on the polar (Zerilli) function.
The transformations between these functions were first found by Chandrasekhar~\cite{Chandrasekhar:1975nkd}. 
For simplicity, we will assume that the primary and secondary effective potentials are unmodified, and focus on the effect of the coupling term exclusively.
The impact of modifications to the effective potential on the ringdown has already been investigated in Refs.~\cite{Thomopoulos:2025nuf,deMedeiros:2025ayq,Nakashi:2025fbr}.
Since our goal is to isolate the effects of mode mixing, it is sufficient to focus on the coupling terms, which mediate the interaction between the primary and secondary wave functions.

To systematically assess the contamination of the gravitational-wave signal by the secondary wave function, we proceed as follows:
\begin{enumerate}
    \item 
    We specify initial data only in the primary field, setting both the secondary field and its initial time derivative to zero. This choice isolates coupling-induced mode excitation: any secondary-field response, and any secondary-led contribution subsequently appearing in the primary signal, is generated dynamically by the coupling rather than inserted through the initial data. This choice also provides the closest analog of a pure GR initial perturbation. In this limited sense, the choice is conservative because it excludes direct excitation of the secondary sector. It should nevertheless be regarded as a benchmark rather than a generic prediction for modified-gravity black hole mergers, where the secondary field may also be excited initially. While we will use Gaussian initial data, we have verified that other types of initial conditions lead to similar numerical results.
    \item We extract both the primary and secondary wave functions far from the peak of the potential, at $x=50M$. We have verified that extracting at larger radii does not significantly change our main results.
    \item We analyze the time-domain signals using two different fitting methods (the Prony method~\cite{Berti:2007dg} and a least-squares fit) to identify the frequencies and amplitudes of the dominant QNMs in the waveform.
    \item We explore the parameter space by repeating the steps above for different values of the coupling parameter $\alpha_j$.
\end{enumerate}

This procedure yields amplitude fits that can be used to estimate the importance of the mixing terms as a function of $\alpha_j$ and to assess if the extra family of modes associated with a given modified theory of gravity may become relevant when compared with subdominant modes or overtones within GR. 

Before getting into the details of our analytical and numerical methods, we will discuss how the modes can be expected to scale in a perturbative analysis where we assume $\alpha_j\ll 1$.
From now on we focus on specific cases where there is only a single power $(1/r)^j$ in the coupling function ${\cal C}(r)$, such that we can remove the subscript $j$ in $\alpha_j=\alpha$, unless stated otherwise.

\subsection{Small-coupling limit: The hierarchical system}\label{sec:hierarquical}

To gain insight into the behavior of the time evolution, it is useful to do a small-$\alpha$ expansion of the differential equations. As stated before, we focus on the case in which a single power of $j$ is present in the coupling. Since we consider an initial pulse in the primary mode and ${\cal C}(r)$ is of order $\alpha$, say ${\cal C}(r)=\alpha \bar{\cal C}(r)$, we can assume the following perturbative structure for the wave functions:
\begin{subequations}
\begin{align}
\Psi_{s_1}&=\Psi_{s_1,(0)}+\alpha^2\,\Psi_{s_1,(2)}+{\cal O}(\alpha^4)\,,\\
\Psi_{s_2}&=\alpha\,\Psi_{s_2,(1)}+{\cal O}(\alpha^3)\,,
\end{align}
\end{subequations}
where the indices in round brackets label the order in $\alpha$.
By substitution into the differential equations we get
\begin{subequations} \label{eq:dPsis1s2s3}
\begin{align}
	\left[ \frac{\del^2 }{\del x^2}- \frac{\del^2 }{\del t^2} - V_{s_1}(r) \right] \Psi_{s_1,(0)}
	&= 0\,.\label{eq:dPsis3}\\
	\left[ \frac{\del^2 }{\del x^2}- \frac{\del^2 }{\del t^2} - V_{s_2}(r) \right] \Psi_{s_2,(1)}
	&= \bar{\cal C}(r)\, \Psi_{s_1,(0)}\,,\label{eq:dPsis2}\\
	\left[ \frac{\del^2 }{\del x^2}- \frac{\del^2 }{\del t^2} - V_{s_1}(r) \right] \Psi_{s_1,(2)}
	&= \bar{\cal C}(r)\, \Psi_{s_2,(1)}\,.\label{eq:dPsis1}
\end{align}
\end{subequations}
Therefore, for $\alpha \ll 1$ the system becomes hierarchical and can be solved order by order. From the structure of the equations, we expect that:
\begin{enumerate}
    \item At leading order, the primary wave function $\Psi_{s_1,(0)}$ is governed by the homogeneous GR equation, Eq.~\eqref{eq:dPsis3}, and hence exhibits the standard GR QNM response.
    \item At first order in $\alpha$, the secondary field is sourced by $\Psi_{s_1,(0)}$ through Eq.~\eqref{eq:dPsis2}. As a result, it contains both a driven response associated with the GR frequencies and the intrinsic family of modes of the secondary field.
    \item At second order in $\alpha$, the primary wave function receives a quadratic correction sourced by $\Psi_{s_2,(1)}$, as shown in Eq.~\eqref{eq:dPsis1}. Since $\Psi_{s_2,(1)}$ carries information about both families of modes, the corrected primary signal will also contain both contributions.
\end{enumerate}
This structure suggests that the
leading coupling dependence of the QNM amplitudes will take the form
\begin{subequations}
\begin{align}
    A_{s_1}&=A_{s_1}^0\left[1+c_{s_1,(2)}\,\alpha^2\right],\label{eq:fit1}\\
    A_{s_2}&=A_{s_1}^0\,c_{s_2,(2)}\,\alpha^2,\label{eq:fit2}
\end{align}
\end{subequations}
when the amplitudes are extracted from the primary-wave signal, and
\begin{subequations}
\begin{align}
    A_{s_1}&=A_{s_1}^0\,c_{s_1,(1)}\,\alpha,\label{eq:fit3}\\
    A_{s_2}&=A_{s_1}^0\,c_{s_2,(1)}\,\alpha,\label{eq:fit4}
\end{align}
\end{subequations}
when they are extracted from the secondary-field signal. Here, the coefficients $c$ are constants, $A_{s_1}^0$ denotes the amplitude associated with the excitation of $\Psi_{s_1,(0)}$ in the GR limit, $A_{\rm s_1}$ is the amplitude of the primary-led mode, and $A_{\rm s_2}$ is the amplitude of the secondary-led mode.
Two remarks are in order. First, the hierarchical system does not capture properly the shifts in the QNM spectrum induced by the coupling between the perturbations, which naturally arise in the full system, as pointed out in Ref.~\cite{McManus:2019ulj} (see also Sec.~\ref{sec:frequencydomain}). Second, the discussion above may be invalid when the system is close to a resonance -- e.g., when the Regge-Wheeler and Zerilli functions are coupled, as in some EFT scenarios~\cite{Cardoso:2018ptl,Cano:2021myl}.

\subsection{Frequency-domain computations}\label{sec:frequencydomain}

From a Fourier expansion of Eqs.~\eqref{eq:Psis1} and~\eqref{eq:Psis2}, namely $\left[\Psi_{s_1},\Psi_{s_2}\right]=\left[\psi(r),\phi(r)\right]e^{-i\omega t}$, we obtain a system of ordinary differential equations which, when supplemented with suitable boundary conditions, defines an eigenvalue problem for the frequency $\omega$. 
The fundamental QNMs of the resulting system of equations was studied 
in Ref.~\cite{McManus:2019ulj}.
A similar study was carried out for the overtones in Refs.~\cite{Volkel:2022aca,Hirano:2024fgp}.
Here, we focus on the fundamental mode and the first overtone of both the primary (gravitational) and secondary master functions.

To compute the modes, we use two complementary methods: direct integration (DI) and the matrix-valued continued-fraction (CF) method. Both techniques are described for coupled systems in Ref.~\cite{Pani:2013pma} and applied, for instance, in Ref.~\cite{Macedo:2016wgh}. The DI method works best for weakly damped modes, as highly damped modes suffer from numerical contamination due to the exponential growth of the wavefunctions at spatial infinity. For this reason, whenever necessary, the results used to fit the time-evolution profiles are obtained with the CF method, while the DI approach is used to validate the fundamental mode frequencies. We describe both methods next.

\subsubsection{Direct integration}

First, we note that both the potential and the coupling function vanish as $x\to\pm\infty$. This allows us to impose the following boundary conditions:
\begin{equation}
    \mathbf{\Psi}=\left\{
    \begin{array}{ll}
        e^{-i \omega x}\sum_{n=0}^{N_{-}}(r-2M)^n \, \mathbf{b}_n\,, & x\to-\infty\,, \\[0.2cm]
        e^{i \omega x}\sum_{n=0}^{N_{+}}r^{-n} \, \mathbf{c}_n\,, & x\to\infty\,,
    \end{array}\right.
\end{equation}
where $\mathbf{\Psi}=(\Psi_{s_1}, \Psi_{s_2})^{T}$ is a column vector. The coefficient vectors $\mathbf{b}_j$ and $\mathbf{c}_j$ are obtained from the frequency-domain version of Eqs.~\eqref{eq:Psis1} and~\eqref{eq:Psis2} by expanding them near the horizon and at infinity, respectively, and solving iteratively for the coefficients of the powers of $(r-2M)$ and $r^{-1}$ up to the desired orders $N_\pm$.
In practice, we use $N_-=6$ and $N_+=10$, and verify the numerical stability of our results under changes in these and other numerical parameters, such as the numerical infinity and horizon, and the matching point (see below). The resulting expansions depend on the two constant vectors $\mathbf{b}_0$ and $\mathbf{c}_0$, which can be conveniently used as a basis for the wave functions.

With these boundary conditions at hand, we can numerically integrate the differential equations. Two linearly independent solutions are constructed by integrating from the horizon and choosing $\mathbf{b}_0=(1,0)$ and $\mathbf{b}_0=(0,1)$, which define $\mathbf{\Psi}^-_{1}$ and $\mathbf{\Psi}^-_{2}$, respectively. Likewise, two independent solutions are obtained by integrating from numerical infinity and choosing $\mathbf{c}_0=(1,0)$ and $\mathbf{c}_0=(0,1)$, yielding $\mathbf{\Psi}^+_1$ and $\mathbf{\Psi}^+_2$, respectively.
A general solution that is purely ingoing at the horizon and purely outgoing at infinity must be a linear combination of these fundamental solutions. To impose the matching conditions, we construct the Wronskian matrix at an intermediate point $x_m$ as
\begin{equation}
\mathbf{X}(x_m)=
\begin{pmatrix}
\Psi_{s1,1}^-& \Psi_{s1,2}^-& \Psi_{s1,1}^+& \Psi_{s1,2}^+\\
\Psi_{s2,1}^-& \Psi_{s2,2}^-& \Psi_{s2,1}^+& \Psi_{s2,2}^+\\
{\Psi_{s1,1}^-}'& {\Psi_{s1,2}^-}'& {\Psi_{s1,1}^+}'& {\Psi_{s1,2}^+}'\\
{\Psi_{s2,1}^-}'& {\Psi_{s2,2}^-}'& {\Psi_{s2,1}^+}'& {\Psi_{s2,2}^+}'
\end{pmatrix}_{x=x_m},
\end{equation}
where $\Psi_{si,j}^{\pm}$ denotes the $i$-th component of the $j$-th independent solution integrated from the horizon ($-$) or from infinity ($+$). The QNM frequencies are then the values of $\omega$ for which
\begin{equation}
    \det\bigl[\mathbf{X}(x_m)\bigr]=0.
\end{equation}
With this procedure, we can compute the fundamental mode of the primary and secondary fields. 

\subsubsection{Matrix-valued continued fraction}

The matrix-valued CF method relies on a Frobenius expansion of the differential equations. An expansion satisfying both boundary conditions can be written as~\cite{Pani:2013pma}
\begin{align}
    \mathbf{\Psi}=e^{i\omega r}\left(\frac{r}{2M}-1\right)^{-2iM\omega}\left(\frac{r}{2M}\right)^{4 iM\omega}\sum_{n=0}^\infty\mathbf{d}_n \, \Bigl(1-\frac{2M}{r}\Bigr)^n\,.
    \nonumber \\
\label{eq:frobenius}
\end{align}
Substituting Eq.~\eqref{eq:frobenius} into the differential equations yields a matrix-valued $n$-term recurrence relation, where $n$ equals the highest power of $2M/r$ retained in the mixing coefficients of Eq.~\eqref{eq:mixing}. Here we focus on the case $n=5$, although the method can be extended to other cases. The resulting recurrence relation has the form
\begin{subequations}
\begin{align}
    &\boldsymbol{\alpha}_0 \mathbf{d}_{1}
    +\boldsymbol{\beta}_0 \mathbf{d}_{0}=0,\\
    &\boldsymbol{\alpha}_1 \mathbf{d}_{2}
    +\boldsymbol{\beta}_1 \mathbf{d}_{1}
    +\boldsymbol{\gamma}_1 \mathbf{d}_{0}=0,\\
    &\boldsymbol{\alpha}_2 \mathbf{d}_{3}
    +\boldsymbol{\beta}_2 \mathbf{d}_{2}
    +\boldsymbol{\gamma}_2 \mathbf{d}_{1}
    +\boldsymbol{\delta}_2 \mathbf{d}_{0}=0,\\
    &\boldsymbol{\alpha}_n \mathbf{d}_{n+1}
    +\boldsymbol{\beta}_n \mathbf{d}_{n}
    +\boldsymbol{\gamma}_n \mathbf{d}_{n-1}
    +\boldsymbol{\delta}_n \mathbf{d}_{n-2}
    +\boldsymbol{\sigma}_n \mathbf{d}_{n-3}=0, \\
    &\textrm{for} \quad n\geq 3, \nonumber
\end{align}
\end{subequations}
where the $2\times2$ matrix coefficients $\boldsymbol{\alpha}_j,\boldsymbol{\beta}_j,\boldsymbol{\gamma}_j,\boldsymbol{\delta}_j$ and $\boldsymbol{\sigma}_j$ are symmetric, with components,
\begin{subequations}
\begin{align}
    \alpha_{kk} &= (n+1)(n-4 i M \omega +1),\quad \alpha_{12}=0, \\
    \beta_{kk}  &=- 2 n^2+n (-2+16 i M \omega )+8 M \omega  (4 M \omega +i)
    \nonumber \\
    &\quad + s_k^2-\ell  (\ell +1)-1,
    \quad \beta_{12}=-\alpha_3-\alpha_4-\alpha_5,\\
    \gamma_{kk} &= (n-4 i M \omega -s_k) (n-4 i M \omega +s_k),
    \\
                \gamma_{12} &= \alpha_3+2 \alpha_4+3 \alpha_5,\\
    \delta_{kk} &= \sigma_{kk}=0,\quad \delta_{12}=-\alpha_4-3 \alpha_5,\quad \sigma_{12}=\alpha_5.
\end{align}
\end{subequations}

This five-term recurrence relation can be reduced to a three-term one by performing two Gaussian elimination steps. As an intermediate step, one first reduces it to a four-term recurrence relation by defining
\begin{subequations}
\begin{align}
    &\tilde{\boldsymbol{\alpha}}_n=\boldsymbol{\alpha}_n,\\
    &\tilde{\boldsymbol{\beta}}_n=\boldsymbol{\beta}_n,\qquad
    \tilde{\boldsymbol{\gamma}}_n=\boldsymbol{\gamma}_n,\qquad
    \tilde{\boldsymbol{\delta}}_n=\boldsymbol{\delta}_n,\qquad n\leq 1,\\
    &\tilde{\boldsymbol{\beta}}_n=\boldsymbol{\beta}_n-\boldsymbol{\sigma}_n[\tilde{\boldsymbol{\delta}}_{n-1}]^{-1}\tilde{\boldsymbol{\alpha}}_{n-1},\qquad n>1,\\
    &\tilde{\boldsymbol{\gamma}}_n=\boldsymbol{\gamma}_n-\boldsymbol{\sigma}_n[\tilde{\boldsymbol{\delta}}_{n-1}]^{-1}\tilde{\boldsymbol{\beta}}_{n-1},\qquad n>1,\\
    &\tilde{\boldsymbol{\delta}}_n=\boldsymbol{\delta}_n-\boldsymbol{\sigma}_n[\tilde{\boldsymbol{\delta}}_{n-1}]^{-1}\tilde{\boldsymbol{\gamma}}_{n-1},\qquad n>1.
\end{align}
\end{subequations}
Repeating the procedure then yields a three-term recurrence relation, which can be solved using the method described in Refs.~\cite{Pani:2013pma,Rosa:2011my}. To this end, we define the ladder operator
\begin{equation}
    \mathbf{d}_{n+1}=\mathbf{R}_n^{+} \, \mathbf{d}_n,
\end{equation}
which, for the three-term recurrence relation, satisfies
\begin{equation}
    \mathbf{R}_n^{+}=-\left[\tilde{\tilde{\boldsymbol{\beta}}}_{n+1}+\tilde{\tilde{\boldsymbol{\alpha}}}_{n+1}\mathbf{R}_{n+1}^{+}\right]^{-1}\tilde{\tilde{\boldsymbol{\gamma}}}_{n+1}.
\label{eq:recurrence}
\end{equation}
Here, double-tilde quantities denote the coefficients of the resulting three-term recurrence relation. Proceeding down to the $j=0$ equation, one finds $\mathbf{M} \, \mathbf{d}_0=0$, where
\begin{equation}
    \mathbf{M}=\tilde{\tilde{\boldsymbol{\beta}}}_{0}-\tilde{\tilde{\boldsymbol{\alpha}}}_{0}\left[\tilde{\tilde{\boldsymbol{\beta}}}_{1}-\tilde{\tilde{\boldsymbol{\alpha}}}_{1}\left(\tilde{\tilde{\boldsymbol{\beta}}}_{2}+\tilde{\tilde{\boldsymbol{\alpha}}}_{2}\mathbf{R}_{2}^{+}\right)^{-1}\tilde{\tilde{\boldsymbol{\gamma}}}_{2}\right]^{-1}\tilde{\tilde{\boldsymbol{\gamma}}}_{1}.
\end{equation}
In practice, we start from a sufficiently large value of $n$, assign a numerical value to $\mathbf{R}_N^{+}$, and then solve Eq.~\eqref{eq:recurrence} backwards to obtain the matrix $\mathbf{M}$. Finally, the QNM frequencies are determined by imposing
\begin{equation}
    \det \mathbf{M}=0.
\end{equation}

With this procedure, we can obtain not only the fundamental modes, but also the first few overtones of each family of modes. The CF method is accurate and stable, and it will be our method of reference in all numerical calculations of QNM frequencies.

\subsection{Time evolution and quasinormal modes}
\label{sec:numerical_methodology_validation}
\subsubsection{Coupled system and solution methods}
\label{subsec:Coup_System_Solution_Methods}
We compute the time-domain response of the coupled system by evolving an initial perturbation with the method of lines~\cite{Berti:2009kk,Pani:2013pma}. In this approach, the spatial coordinate is discretized while time remains continuous, reducing the system to a set of ordinary differential equations that can be integrated numerically.
The evolution is performed in the tortoise coordinate $x$, defined through the mapping $r=r(x)$, which maps the exterior domain into $x\in(-\infty,\infty)$ and allows for an accurate numerical treatment of both the near-horizon and asymptotic regions.

As our initial data, we impose a Gaussian profile in the gravitational sector,
\begin{equation}
	\Psi_{s_1}(x, t_0) = A e^{-b(x - x_0)^2/M^2},\label{eq:gaussian}
\end{equation}
while the additional field is initially set to $\Psi_{s_2}(x,t_0)=0$. The initial time derivatives of both fields are taken to vanish,
\begin{equation}
	\left.\frac{\partial \Psi_{s_i}}{\partial t}\right|_{t=t_0} = 0\,, \quad i=1,\,2\,.
\end{equation}

The spatial domain in the tortoise coordinate, $x \in [x_{\rm min},x_{\rm max}]$, is uniformly discretized as $x_i = x_{\rm min} + i h$, where $h$ is the grid spacing and $i=0,\dots,N$. Second-order spatial derivatives are approximated with central finite differences,
\begin{equation}
	\left.\frac{\del^2\mathbf{\Psi}}{\del x^2}\right|_{x=x_i} \approx \frac{\mathbf{\Psi}_{i-1} - 2\mathbf{\Psi}_i + \mathbf{\Psi}_{i+1}}{h^2}.
\end{equation}
The resulting system of coupled ordinary differential equations is integrated in time using the \texttt{NDSolve} routine in \texttt{Mathematica}. As usual in time-domain analyses, the waveform is extracted at a fixed observation point, and its temporal evolution is then monitored.
To avoid contamination from the boundaries, namely the near-horizon and asymptotic regions, the numerical domain is chosen such that these regions remain causally disconnected from the extraction point during the time interval of interest~\cite{Allen:2004js}.
\begin{figure*}
	\centering
	\centering
	\includegraphics[width=0.48\textwidth]{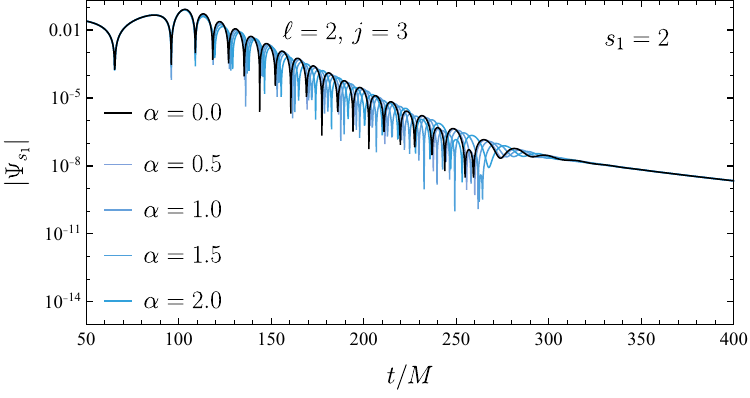}
	\includegraphics[width=0.48\textwidth]{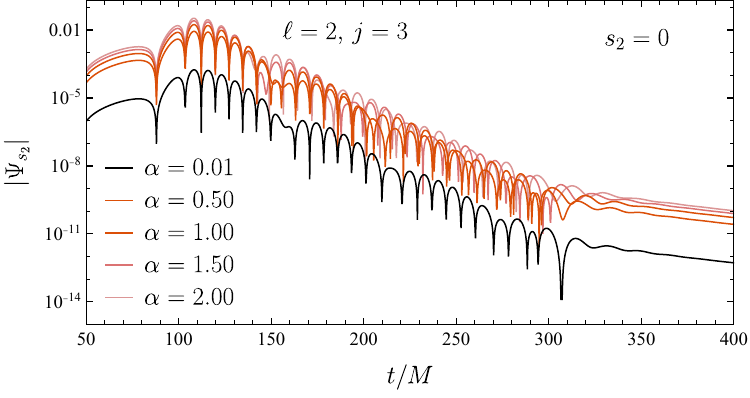}\\
    \includegraphics[width=0.48\textwidth]{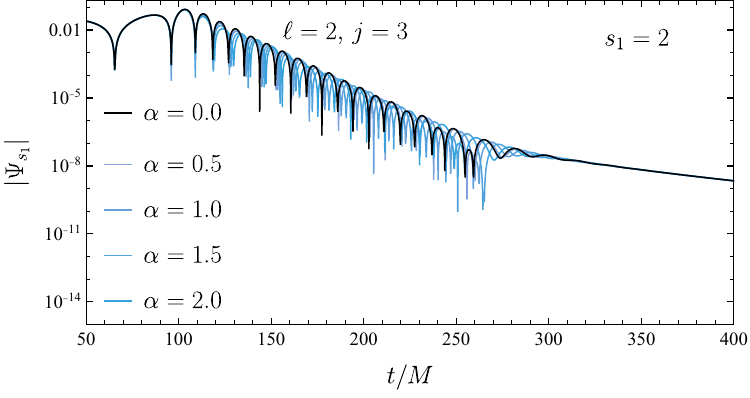}
	\includegraphics[width=0.48\textwidth]{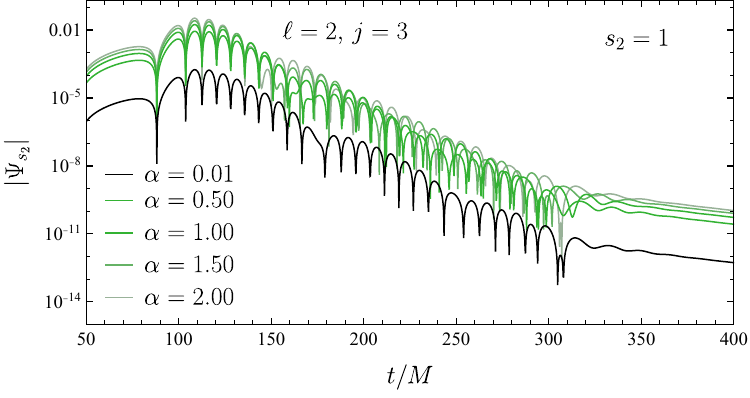}
	\caption{Time-domain evolution of the field amplitudes $|\Psi|$ as functions of $t/M$ for coupled systems with $\ell = 2$ and $j = 3$.
	The top panels show the gravitational--scalar coupling: gravitational component (left) and scalar component (right).
	The bottom panels show the gravitational--vector coupling: gravitational component (left) and vector component (right).
	All panels display evolutions for selected coupling strengths $\alpha = 0.5$, $1.0$, $1.5$, and $2.0$. The left panels also show the decoupled GR limit $\alpha = 0$ (in black), while the right panels include the weak-coupling case $\alpha = 0.01$.}
	\label{fig:timeevolution05_j3}
\end{figure*}

To illustrate the evolution of the coupled system, in Fig.~\ref{fig:timeevolution05_j3} we present the time-domain profiles obtained for $\ell=2$ and $j=3$, for different values of the coupling strength $\alpha$. The initial data consist of a Gaussian pulse in the gravitational sector, with parameters $A=1$, $b=1$, and $x_0=50M$, while the additional field ($s_2=0,\,1$) is initially set to zero.
The signals exhibit the standard structure of black hole perturbations, which can be divided into three stages: (i) an initial transient, or prompt response, whose morphology depends strongly on the chosen initial data; (ii) the ringdown phase, characterized by damped oscillations governed by the QNMs, with excitation amplitudes and phases that also retain an imprint of the initial data; and (iii) a late-time tail, in which the signal amplitude decays significantly and approaches a quasi-stationary regime.
However, in contrast to the uncoupled case, the interaction between the sectors leads to the excitation of both fields, even though only the gravitational sector is initially present. This clearly demonstrates the presence of mode mixing and energy transfer induced by the coupling.

\subsection{Extraction of quasinormal modes: Prony method}

The QNM spectrum of the coupled system can be extracted from our waveform during the ringdown. To do so, we identify a suitable ringdown window by initially taking the starting time to coincide with the peak of the waveform, $t_{\rm start}=t_{\rm peak}$, while keeping the ending time $t_{\rm end}$ fixed. We take  $t_{\rm start}=t_{\rm peak}$
only as a lower bound for the fitting window, rather than identifying the waveform peak with a unique physical ringdown onset. Indeed, recent spectral decompositions indicate that the transition from the prompt response to a regime dominated by the fundamental mode and low overtones can extend beyond the first waveform peak \cite{Besson:2026aog}. Therefore, we increase $t_{\rm start}$ until the extracted frequencies and amplitudes reach a sufficiently stable regime. We fix $t_{\rm end} = t_{\rm peak} + 180M$. We assessed the stability of our results by shifting $t_{\rm end}$ by $\pm 20M$ and we find that the extracted mode frequencies and amplitudes do not change appreciably under this variation.

For a given field $\Psi(t)$, the ringdown signal can be modeled as a sum of complex exponentials,
\begin{equation}
	\Psi(t) \simeq \sum_{k=1}^{p} C_k e^{-i \omega_k (t-t_{\rm peak})},
	\label{eq:prony_cont}
\end{equation}
where
$t_{\rm peak}$ is the time at the peak of the waveform,
$\omega_k$ are the complex quasinormal frequencies, and $C_k$ are complex amplitudes ($C_k$ encodes both the amplitude and the phase of the modes, i.e., $C_k=A_k e^{i \phi_k}$). Notice that the primary and secondary signals have different values of $t_{\rm peak}$, with the secondary peaking approximately $5 M$ after the primary (see Fig.~\ref{fig:timeevolution05_j3}).
Sampling the signal at discrete times $t_n = n \Delta t$, Eq.~\eqref{eq:prony_cont} can be rewritten as
\begin{equation}
	x_n = \Psi(t_n) = \sum_{k=1}^{p} C_k z_k^n,
	\label{eq:prony_disc}
\end{equation}
where $z_k = e^{-i \omega_k \Delta t}$. The problem of extracting the QNM frequencies is therefore reduced to determining the complex parameters $z_k$ from the discrete data $x_n$.

To achieve this, we use a Prony-type method in its modern formulation based on singular value decomposition (SVD); see e.g., Ref.~\cite{Berti:2007dg} for an early application of Prony methods to the analysis of  ringdown waveforms, and Ref.~\cite{Geissler:2026iri} for a more recent discussion. In this method, the time series is first restricted to a suitable interval within the ringdown phase. A Hankel matrix is then constructed from the data, and its singular values are computed via SVD. The number of relevant modes, $p$, is determined by retaining the dominant singular values above a prescribed tolerance, thereby filtering out numerical noise. In practice, $p$ must be chosen large enough to capture the dominant modes present in the signal, but not so large that noise-dominated components are incorrectly interpreted as physical modes. In our cases, we find that even for relatively large values of $p$ (say, $p=100$), the extracted frequencies remain consistent provided that only the numerically stable modes are ultimately retained.

Once $p$ is fixed, the coefficients of the characteristic polynomial
\begin{equation}
	A(z) = \prod_{k=1}^{p} (z - z_k) = \sum_{m=0}^{p} \alpha_m z^{p-m},
	\label{eq:poly}
\end{equation}
are found by solving a linear system of equations derived from Eq.~\eqref{eq:prony_disc}. This system is solved using a pseudo-inverse, which improves numerical stability in the presence of noisy data.
The roots $z_k$ of Eq.~\eqref{eq:poly} are then computed numerically, and the corresponding quasinormal frequencies are recovered through
\begin{equation}
	\omega_k = i \ln(z_k)/\Delta t.
	\label{eq:freqs}
\end{equation}
Finally, the amplitudes $C_k$ are determined by solving a linear system relating the extracted modes to the original data. This allows for a full reconstruction of the ringdown signal in terms of its QNM content.

In coupled systems, the interaction between sectors leads to the simultaneous excitation of multiple modes in each field. As a consequence, both the gravitational and the additional fields contain contributions from different branches of the spectrum, reflecting the presence of mode mixing. The SVD-based Prony method is particularly well suited for this analysis, as it enables a robust identification of multiple damped modes without requiring prior knowledge of the frequencies.

As a complementary approach to extract the QNM frequencies from the time-domain data, we also employ a nonlinear optimization procedure, treating the frequencies directly as adjustable parameters. This approach is especially valuable when physical effects such as mode mixing or parametrized deformations of the potential shift the frequencies away from their standard GR values~\cite{Berti:2007dg,Liang:2026eic,Yang:2022ifo,Baibhav:2023clw,Cheung:2023vki}.

\subsection{Nonlinear least-squares extraction of amplitudes and frequencies}
As a second method to determine the QNM content directly of the numerical waveforms, we use a nonlinear least-squares fitting procedure in which both the amplitudes and the complex frequencies are treated as free parameters. The ringdown signal is modeled as a superposition of damped oscillations,
\begin{equation}
	h(t) = \sum_{k=1}^{N} A_k \, e^{-\omega_{I,k} (t-t_{\rm peak})} \, \cos[\omega_{R,k} (t-t_{\rm peak}) + \phi_k],
	\label{eq:ringdown_model}
\end{equation}
where 
$N$ denotes the number of modes included in the fit. In our analysis, we set $N=2$, using the fundamental gravitational-led and scalar-led modes as initial guesses, and  
we denote by $\omega_k = \omega_{R,k}-i\omega_{I,k}$ the complex QNM frequency of the $k$th mode. In this parametrization, the fitting parameters are the amplitudes $A_k$, phases $\phi_k$, oscillation frequencies $\omega_{R,k}$, and damping rates $\omega_{I,k}$.

The amplitudes, phases and QNM frequencies can be inferred directly from the waveform by minimizing the quadratic residual between the numerical signal and the fitted model~\eqref{eq:ringdown_model}. Given a set of sampled data $\{t_i, h_i\}$, we minimize
\begin{equation}
	\chi^2 = \sum_i \left[h_i - h_{\text{fit}}(t_i)\right]^2,
	\label{eq:minimize}
\end{equation}
where $h_{\text{fit}}(t_i)$ is given by Eq.~\eqref{eq:ringdown_model} at discrete times $t = t_i$.

Since the frequencies appear inside the exponential and oscillatory functions in Eq.~\eqref{eq:ringdown_model}, the fitting problem is intrinsically nonlinear; see e.g., Ref.~\cite{Giesler:2024hcr}.
Consequently, the minimization of Eq.~\eqref{eq:minimize} must be performed iteratively using nonlinear optimization algorithms. In practice, the procedure requires suitable initial guesses for the fitting parameters in order to ensure convergence toward physically relevant solutions. When it converges, the method provides simultaneous estimates for the oscillation frequencies $\omega_{R,k}$, damping rates $\omega_{I,k}$, amplitudes $A_k$, and phases $\phi_k$.
In such cases, allowing the frequencies to vary freely enables the identification of mode shifts, mode mixing, and additional spectral components that may not be captured by a fixed-frequency analysis.
Yet, nonlinear fitting procedures are generally more computationally expensive and sensitive to the choice of fitting window and initial parameters. Moreover, the presence of multiple damped oscillatory components may lead to parameter degeneracies and local minima in the optimization landscape. For this reason, nonlinear fits are often complemented by independent spectral analyses or by comparisons with theoretically predicted QNM frequencies.

\begin{figure*}
    \includegraphics[width=\linewidth]{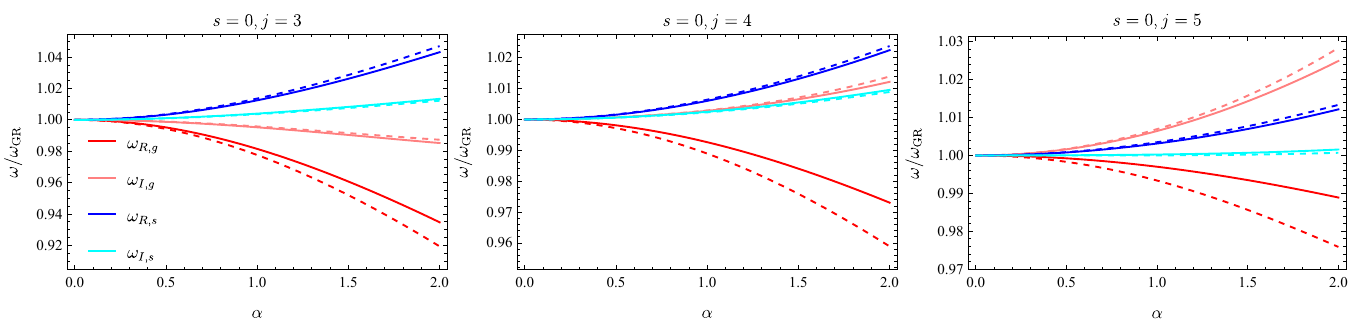}\\
    \includegraphics[width=\linewidth]{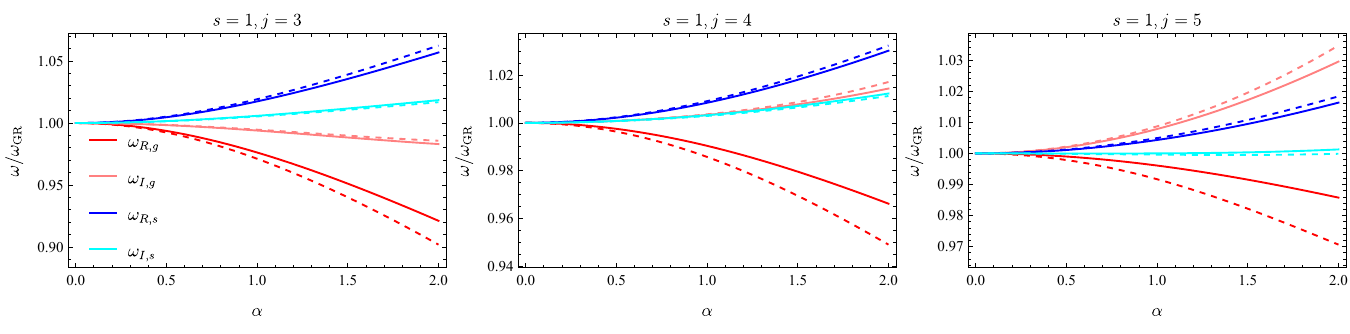}
    \caption{Quasinormal frequencies computed with the CF method, normalized by their GR values. We show the fundamental mode (solid lines) and the first overtone (dashed lines). The coupling affects both the fundamental mode and the first overtone in a qualitatively similar way in all cases considered, and the deviations from the GR values scale quadratically with the coupling.}
    \label{fig:qnm_cf}
\end{figure*}

\section{Numerical results}\label{sec:results}

\subsection{Quasinormal mode frequencies}

We use both the DI and CF methods described in Sec.~\ref{sec:frequencydomain}. In general, because of the radial behavior of the QNMs, the DI method is not well suited for extracting overtones. Therefore, we use it mainly to validate the CF method before exploring the parameter space (in  Appendix~\ref{app:frequencies} we compare the results obtained using the two methods).
We then compute the fundamental mode and the first overtone of the system described by Eqs.~\eqref{eq:Psis1} and \eqref{eq:Psis2} with the CF method. The results are shown in Fig.~\ref{fig:qnm_cf}. Within the parameter range considered here, the shifts in both the fundamental mode and the first overtone scale quadratically with the coupling, in agreement with the behavior reported in Refs.~\cite{McManus:2019ulj,Volkel:2022aca,Hirano:2024fgp}. Deviations from GR are smaller than $10\%$ within the interval $\alpha\in[0,2]$. In the next section, we compare these frequency-domain results with the frequencies extracted from the time-domain evolution.

\subsection{Theory-agnostic cases}

We now study the time-domain evolution of the coupled system. As we have seen by comparing the left and right panels of Fig.~\ref{fig:timeevolution05_j3}, mode mixing implies that the initially excited gravitational sector sources oscillations in the scalar and vector fields via their coupling. As $\alpha$ grows, the secondary field amplitude increases, and so do modifications in the gravitational waveforms. This indicates larger deviations in the relative excitations, reflecting the impact of the coupling on the ringdown dynamics. In our study, we consider different values of $j\in[3,15]$, $\alpha\in[0,2]$, and the spins of the secondary field, $s_2 = 0$, 1, and 2. We then analyze the impact of the coupling $\alpha$ on the signals by extracting the QNM amplitudes and frequencies.

\begin{figure*}
    \includegraphics[width=0.9\linewidth]{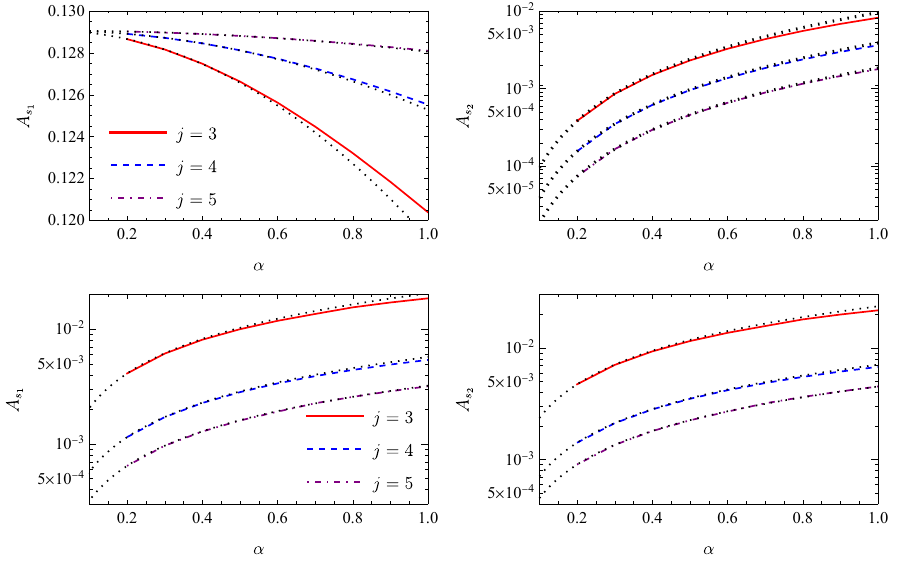}
    \caption{QNM amplitudes as a function of the coupling $\alpha$, for different values of $j$, in the case where the secondary field is a scalar. The upper panels refer to the amplitudes of the gravitational signal, and the bottom panels to those of the scalar signal. The dotted lines correspond to analytical fits with a dependence $\sim \alpha^2$ for the gravitational signal, and $\sim \alpha$ for the scalar signal.}
    \label{fig:amp_prony_scalar}
\end{figure*}

\begin{figure*}
    \includegraphics[width=0.9\linewidth]{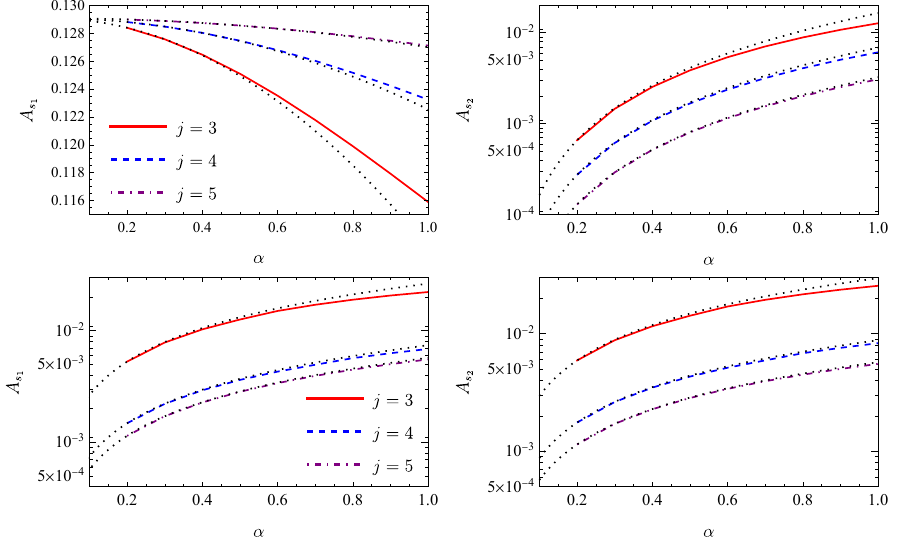}
    \caption{QNM amplitudes as a function of the coupling $\alpha$, for different values of $j$, in the case where $s_2=1$ (i.e., the secondary field is a vector). The upper panels refer to the amplitudes of the gravitational signal, and the bottom panels to those of the vector signal. The dotted lines correspond to analytical fits with a dependence $\sim \alpha^2$ for the gravitational signal, and $\sim \alpha$ for the vector signal.}
    \label{fig:amp_prony_vector}
\end{figure*}

From the time-evolution signals, we can extract the frequencies and amplitudes using either the Prony method or a least-squares fit. We choose the end of the ringdown window to lie close to the point at which the oscillatory behavior ceases, and we verify that the results are insensitive to variations of this endpoint by $\pm 20M$. In Appendix~\ref{app:Benchmarking}, we present a comparison between the Prony and least-squares methods. Since the Prony method yields the most stable numerical results over a wider range of initial times, we adopt it as our method of choice for extracting amplitudes and frequencies.

Because the perturbative expansion of the frequencies for small $\alpha$ was studied in various papers~\cite{McManus:2019ulj,Volkel:2022aca,Hirano:2024fgp}, we focus on the dependence of the QNM amplitudes on $\alpha$. The results are shown in Figs.~\ref{fig:amp_prony_scalar} and~\ref{fig:amp_prony_vector}, where we present the cases $s_2=0$ and $s_2=1$, respectively, for $j=3,\,4$ and $5$.

The results are qualitatively similar for both choices of the spin of the secondary field, and they are consistent with the perturbative analysis of Sec.~\ref{sec:hierarquical}. From the gravitational-wave signal we find that the ringdown can indeed be decomposed into two contributions --- a gravitational-led mode and a secondary-led mode --- whose amplitudes are well described, in the small-coupling limit (see the dotted lines in Figs.~\ref{fig:amp_prony_scalar} and~\ref{fig:amp_prony_vector}), by the relations given by Eqs.~\eqref{eq:fit1} and \eqref{eq:fit2}.
On the other hand, the amplitudes extracted from the secondary-field signal follow, in the small-coupling limit, the behavior given by Eqs.~\eqref{eq:fit3} and~\eqref{eq:fit4}.  Note that the same symbols $A_{s_1}$ and $A_{s_2}$ are used in Eqs.~\eqref{eq:fit1}--\eqref{eq:fit2} and~\eqref{eq:fit3}--\eqref{eq:fit4} to denote the amplitudes of the same mode families as measured, respectively, in the primary and in the secondary signal, but these are distinct quantities.

\begin{figure*}
    \centering
    \includegraphics[width=0.33\linewidth]{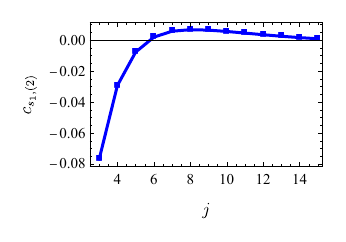}\includegraphics[width=0.33\linewidth]{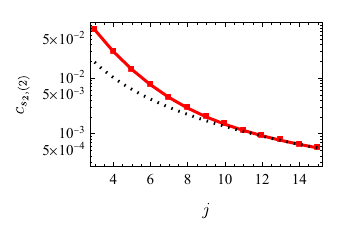}\includegraphics[width=0.33\linewidth]{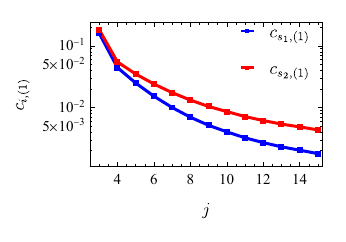}
    \caption{Values for the coefficients up to $j=15$ extracted from the amplitudes of the primary mode (left panel) and of the secondary mode (middle panel) in the primary sector. In the right panel we show the coefficients extracted from the amplitudes of the secondary channels. The behavior of the scalar mode presented in the signal scales approximately as $\propto1/j^{2}$ for large $j$ (dotted line in the central panel).
    }
    \label{fig:coefs}
\end{figure*}

We carried out the same experiment for values of $j$ up to $15$. In Table~\ref{tab:coefficients}, we summarize the values of these coefficients for different choices of $j$ and $s$.
We also ran a few different simulations, changing the values of $(A,b,x_0)$ in the initial condition~\eqref{eq:gaussian} within the ranges $A \in [0.5,2]$, $b=(0.5,2)$, and $x_0=(30,50)$. We found that the relative differences in the coefficients are at most of ${\cal O}(1\%)$. Hence, we keep our original initial data as a reference, and assume that these coefficients can be extrapolated to estimate, at least in order of magnitude, the effects of additional fields coupled to gravity.
As illustrated in Fig.~\ref{fig:coefs}, we see that $c_{s_1,(2)}$ becomes positive when $j\geq6$ for scalar fields, and when $j\geq7$ for vector fields. Curiously, a similar sign change also happens for the corrections to the QNM frequencies~\cite{McManus:2019ulj}. 

\begin{table*}
     \caption{Coefficients entering the small-coupling expansion of the QNM amplitudes as a function of $\alpha$, for different values of $j$. In this case, we use Gaussian initial data with $(A,b,x_0)=(1,1,50M)$. We have also explored other choices in the range $A \in [0.5, 2]$, $b \in [0.5, 2]$, and $x_0 \in [30, 50]M$, and found that the corresponding coefficients agree at the percent level. For reference, the GR value of the gravitational-led amplitude is $A_{s_1}^0\approx 1.29\times 10^{-1}$.}
    \begin{ruledtabular}
    \begin{tabular}{ c | c  c  c  c | c  c  c  c}
        $j$ & \multicolumn{4}{c|}{$s_2=0$} & \multicolumn{4}{c}{$s_2=1$} \\
         &$c_{s_1,(2)}$ & $c_{s_2,(2)}$ & $c_{s_1,(1)}$ & $c_{s_2,(1)}$ & $c_{s_1,(2)}$ & $c_{s_2,(2)}$ & $c_{s_1,(1)}$ & $c_{s_2,(1)}$ \\
         \hline
 3 & -0.0768229 & 0.0739657 & 0.159344 & 0.182793 & -0.127414 & 0.126851 & 0.204209 & 0.228771 \\
 4 & -0.0293212 & 0.0300258 & 0.0443717 & 0.0547823 & -0.0504885 & 0.0528926 & 0.0572678 & 0.0678916 \\
 5 & -0.00779757 & 0.014236 & 0.024943 & 0.0349232 & -0.0160309 & 0.0249967 & 0.0336889 & 0.0441814 \\
 6 & 0.00186873 & 0.00757107 & 0.0152595 & 0.0240349 & -0.000791886 & 0.013162 & 0.0210957 & 0.0303602 \\
 7 & 0.00567285 & 0.0044582 & 0.00997836 & 0.0174051 & 0.00541024 & 0.00771935 & 0.0138493 & 0.0215597 \\
 8 & 0.0066742 & 0.00287133 & 0.00695396 & 0.0131827 & 0.00747364 & 0.00495815 & 0.00984899 & 0.0163482 \\
 9 & 0.00638487 & 0.0019947 & 0.00512468 & 0.0103677 & 0.00729219 & 0.00351876 & 0.00739044 & 0.0128887 \\
 10 & 0.00552921 & 0.00147172 & 0.00402355 & 0.00856847 & 0.00651858 & 0.00264414 & 0.00589312 & 0.0106814 \\
 11 & 0.00448301 & 0.00113906 & 0.00321543 & 0.00711242 & 0.00534927 & 0.00206482 & 0.0047712 & 0.00889569 \\
 12 & 0.0034206 & 0.000914561 & 0.00268789 & 0.00613139 & 0.004023 & 0.00174692 & 0.0040319 & 0.00769093 \\
 13 & 0.00243412 & 0.000756404 & 0.00229428 & 0.00536261 & 0.00280285 & 0.00151864 & 0.00347282 & 0.0067449 \\
 14 & 0.00155736 & 0.000639546 & 0.00202604 & 0.00483692 & 0.00181075 & 0.00127294 & 0.0030903 & 0.00609677 \\
 15 & 0.000798441 & 0.000551887 & 0.00178023 & 0.00431864 & 0.000820778 & 0.00111833 & 0.00273306 & 0.00545592 \\
    \end{tabular}
     \label{tab:coefficients}
     \end{ruledtabular}
\end{table*}

One may also ask how the results would change if, instead of the Regge-Wheeler equation for the primary, we considered the Zerilli equation coupled to additional fields. To address this point, we perform time-domain evolutions considering the Zerilli-scalar system using a Gaussian initial profile with $(A,b,x_0)=(1,1,50)$ and we find results similar to those reported in Table~\ref{tab:coefficients}, with the coefficients showing differences at the level of ${\cal O}(10\%)$. Notably, this difference is an order of magnitude larger than the variations induced by changing the initial data within reasonable ranges, ${\cal O}(1\%)$, indicating that the fitted coefficients are more sensitive to the choice of the primary sector than to the details of the initial configuration. Since our main goal is to estimate the order of magnitude of the excitation of the additional modes during ringdown, we conclude that the analysis based on the Regge-Wheeler function is sufficient for this purpose.

\subsection{Dynamical Chern-Simons gravity}
Dynamical Chern-Simons gravity is an extension of GR in which gravity is coupled to a dynamical scalar field through a parity-violating curvature term (see Ref.~\cite{Alexander:2009tp} for a review). The theory is motivated by effective-field-theory and high-energy considerations, and it is often used as a framework to test possible deviations from GR in gravitational-wave and black hole observations.

Spherically symmetric black holes in dCS gravity are described by the Schwarzschild solution. Reference~\cite{Cardoso:2009pk} showed that the axial sector couples to the scalar field according to
\begin{subequations} \label{eq:dcs}
    \begin{align} 
      V_{11}&=V_2,~V_{22}=V_{0}+f(r)\frac{1}{r_H^2}\frac{144\pi \ell(\ell+1)}{\beta r_H^4}\left(\frac{r_H}{r}\right)^8,\\
      {\cal C}(r)&=f(r)\frac{1}{r_H^2}\frac{12}{\sqrt{\beta}\,r_H^2}\sqrt{\pi\,\frac{(\ell+2)!}{(\ell-2)!}}\left(\frac{r_H}{r}\right)^5,
    \end{align}
\end{subequations}
where $r_H=2M$ and $\beta$ is the dCS coupling parameter.
Note that the GR limit corresponds to $\beta\to\infty$ or $\alpha\to 0$ in the parametrized framework. In this limit, both the scalar perturbation sourced by the Chern–Simons term and the coupling $\mathcal{C}(r)\propto\beta^{-1/2}$ vanish. The opposite limit, $\beta\to 0$,  corresponds to infinitely strong coupling and falls outside the regime of validity of the perturbative expansion used here, which requires $\beta M^4\gg 1$.
(Note that our conventions differ slightly from those of Refs.~\cite{Molina:2010fb,Cardoso:2009pk}; see also Ref.~\cite{McManus:2019ulj}).
The structure of Eqs.~\eqref{eq:dcs} is  similar to the system described by Eqs.~\eqref{eq:Psis1} and~\eqref{eq:Psis2}, with the identification $j=5$ and
\begin{equation}
    \alpha = \frac{3\sqrt{\pi}}{\sqrt{\beta}M^2} \,
    \sqrt{\frac{(\ell+2)!}{(\ell-2)!}}\,,
\end{equation}
where $\beta$ has dimensions of $L^{-4}$. Using the numerical fits from Table~\ref{tab:coefficients}, we can then obtain the corresponding expressions for the mode amplitudes in dCS gravity. Explicitly, for the $\ell=2$ mode, the amplitudes are
\begin{equation}
    A_{\rm s_1}=A_{s_1}^0\left(1-\frac{5.29}{\beta M^4}\right),
    \quad
    {\rm and}
    \quad
    A_{\rm s_2}=A_{s_1}^0\frac{9.66}{\beta M^4}.
\end{equation}
 This shows that, in the weak-coupling regime, where $\beta M^4 \gg 1$, the secondary mode is suppressed, but it still carries a distinctive signature of the parity-violating coupling.

We neglect the effects of the parametric perturbation of the diagonal term on the amplitude. Although this perturbation can produce appreciable shifts in the mode frequency, the corresponding amplitude changes appear to be small when the perturbation scales as $1/r^8$, as in dCS gravity (see Ref.~\cite{Thomopoulos:2025nuf} for more details).

\subsection{Zerilli/Regge-Wheeler system: The degenerate case}

We now consider a coupled system in which the primary field is described by the Zerilli function, while the secondary field is given by the Regge-Wheeler function.
Unlike the previous cases, this system is particularly interesting because the two wave equations are \textit{isospectral} when decoupled. This suggests the possibility of resonances, since the initial perturbation in the primary field acts as a source for the secondary one. As a consequence, the low-$\alpha$ expansion for the amplitude of the secondary mode in the gravitational-wave response, carried out in Sec.~\ref{sec:hierarquical}, becomes invalid. We hence rely here on a fully numerical analysis of the system.

As before, we analyze the time evolutions using Gaussian initial data for different values of $j$ and $\alpha$. The dominant frequencies and amplitudes are extracted with the Prony method, and we verify that the mode frequencies agree with those obtained through DI for the fundamental mode.

\begin{figure}
    \centering
    \includegraphics[width=\columnwidth]{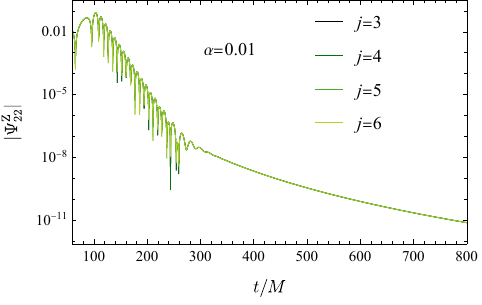}
    \includegraphics[width=\columnwidth]{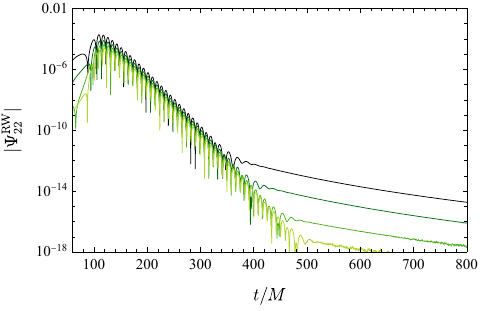}
    \caption{Ringdown waveforms for the coupled Zerilli/Regge-Wheeler system with $\alpha=0.01$. Top panel: Zerilli wavefunction extracted at $50M$ as a function of time. Bottom panel: Regge-Wheeler wavefunction extracted at $50M$. The delay in the onset of the late-time tail increases with $j$, leading to a longer ringdown window in the Regge-Wheeler sector for higher $j$.}
    \label{fig:waveforms_grav}
\end{figure}

In Fig.~\ref{fig:waveforms_grav}, we show the waveforms generated by Gaussian scattering in the coupled Zerilli/Regge-Wheeler system, with $j\in [3,\,6]$ and for fixed $\alpha=0.01$. We see that the onset of the late-time tail for the RW function is delayed, with the delay depending on $j$: larger values of $j$ lead to a later appearance of the tail. As a result, the ringdown lasts longer for the secondary field. We focus on the case $\alpha=0.01$ to emphasize that this effect in the secondary (Regge-Wheeler) channel remains clearly visible even for very small values of the coupling. We have also verified that, for $\alpha=1$, the extended ringdown window in the secondary channel persists, changing slightly, the main difference being in the overall amplitude scale. The in-depth analysis of this result will be left for future work. Here we focus on the QNM amplitudes and frequencies.

\begin{figure}
    \centering
    \includegraphics[width=1\linewidth]{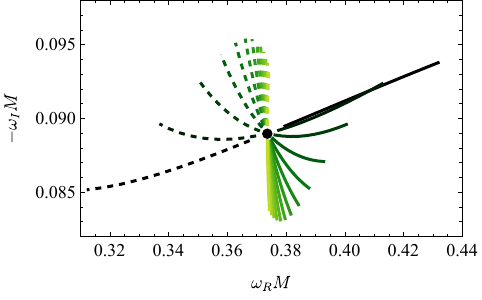}
    \caption{Quasinormal frequencies of the coupled Zerilli/Regge-Wheeler system extracted from the time-domain evolutions, considering a range of values from $j=3$ (darker colors) to $j=15$ (lighter colors), and values of $\alpha$ from $0.2$ to $2$. The black dot marks the decoupled Schwarzschild limit, where the Regge-Wheeler and Zerilli modes are degenerate. The coupling breaks this degeneracy, leading to two distinct families of modes: for $j=15$, the branch along which $|\omega_I| M$ increases with increasing $\alpha$ corresponds to the Regge-Wheeler modes (dashed lines), while the branch along which $|\omega_I| M$ decreases corresponds to the Zerilli modes.}
    \label{fig:rwz}
\end{figure}

In Fig.~\ref{fig:rwz}, we show the mode frequencies extracted with the Prony method from the time-domain signal, considering $j\in[3,15]$ (from darker to lighter colors) and $\alpha\in[0.2,2]$, with the modes departing from the GR limit, represented by the central marker. The modes are extracted from the secondary channel, where the values seem to be  numerically more stable and they agree with the DI results.
The presence of the coupling term breaks the degeneracy of the decoupled system, producing two distinct families of modes. For large values of $j$, the real part deviates only slightly from the GR reference value, indicated by the central black marker, while the most significant changes appear in the imaginary part. For large values of $j$, the imaginary part of the Zerilli QNM frequencies increases.

\begin{figure}
    \centering
    \includegraphics[width=0.95\linewidth]{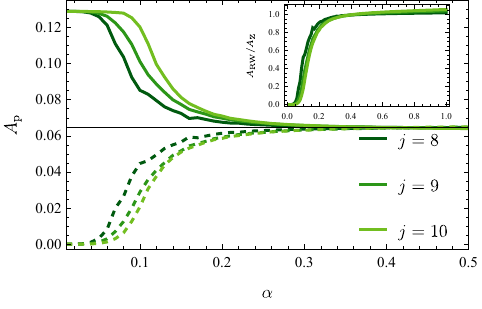}\\
    \includegraphics[width=0.95\linewidth]{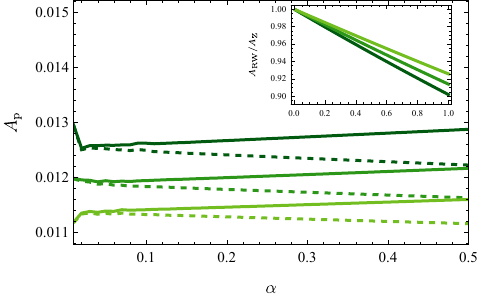}
  \caption{Amplitudes for the coupled Zerilli (solid lines)/Regge-Wheeler (dashed lines) system, considering $j=8,9,$ and $10$. Top panel: Amplitudes extracted from the Zerilli wave function. {In the plot, the subscript ``$\rm p$'' stands for either $\rm Z$ or $\rm RW$}. The amplitude associated with the Regge-Wheeler mode vanishes as $\alpha\to0$, then grows rapidly for moderate values of $\alpha$, saturating at roughly half of the GR amplitude. The inset shows the ratio between the amplitudes, $A_{\rm RW}/A_{\rm Z}$. Bottom panel: Same as in the top panel, but for the amplitudes extracted from the Regge-Wheeler wave function. }
    \label{fig:amplitude}
\end{figure}

For the amplitudes, to explore the small-$\alpha$ regime, we examine in more detail the interval $\alpha\in[0.01,0.5]$, refining the numerical grid resolution to investigate possible changes in behavior relative to larger values of the coupling. We focus on the cases $j=8$, 9, and 10, although the qualitative behavior is the same for other values of $j$. The results for the Zerilli function are shown in the left panel of Fig.~\ref{fig:amplitude}. For $\alpha\approx 0$, the amplitude of the secondary mode starts from $A_{\rm s_2}=0$ and rapidly grows with $\alpha$ until it approaches $\approx A_{s_1}^0/2$ (horizontal line in the plot).
This growth is fast and does not follow the polynomial behavior expected in nondegenerate systems, as far as we can tell numerically (for low values of $\alpha$ it is hard to find the mode decomposition, as the two mode amplitudes are very different from each other). The growth is even faster for lower $j$, as the coupling is stronger. We also find that, because the frequencies of the two modes are very close to each other in this regime, extracting them reliably for small $\alpha$ becomes challenging. In practice, we have to substantially increase the order used in the Prony method to obtain stable fits.

We now turn to the mode mixing in the signal of the secondary (Regge-Wheeler) field. In the top panel of Fig.~\ref{fig:amplitude} we show the amplitudes as a function of the coupling parameter. 
The Regge-Wheeler signal (bottom panel) indicates that the two modes are excited with nearly equal amplitudes for small values of $\alpha$, and then gradually separate as the coupling increases.
We attribute the kink at very small $\alpha$ to numerical difficulties in resolving the amplitudes of the two nearly degenerate modes.
We find that the ratio of the amplitudes, $A^{\rm RW}/A^{\rm Z}$, scales linearly with $\alpha$, as shown in the inset of the left panel of Fig.~\ref{fig:amplitude}, with a slope that depends on $j$. However, 
even for relatively large values of $\alpha$, the difference between the amplitudes remains small.

From an observational perspective, this raises the question of whether such degenerate perturbations could leave identifiable imprints in the gravitational-wave signal. At large distances from the BH, assuming that the GR limit is recovered, the Regge-Wheeler and Zerilli functions contribute to the gravitational-wave polarizations according to
\begin{equation}
    h_+ - i h_\times
    =
    \frac{1}{2r}\sum_{\ell m}
    \sqrt{\frac{(\ell+2)!}{(\ell-2)!}}
    \left(
    \Psi^{\rm Z}_{\ell m}
    +
    i \Psi^{\rm RW}_{\ell m}
    \right)
    {}_{-2}Y_{\ell m},
\end{equation}
where ${}_{-2}Y_{\ell m}$ are the spin-weighted spherical harmonics~\cite{Berti:2009kk}. Therefore, the observed strain contains a combination of the even- and odd-parity sectors, rather than each contribution separately. In principle, if one had access to the full perturbation data or to a multipolar decomposition of the waveform, the two sectors can be disentangled. However, from a single strain signal alone this separation is, in general, not unique. This is particularly relevant in the Schwarzschild case, for which the Regge-Wheeler and Zerilli sectors are isospectral, so that their frequencies alone do not allow one to distinguish between them.
{Nevertheless, isospectrality does not imply identical time-domain excitation. Besson and Jaramillo showed that the Zerilli response can contain an intermediate phase controlled by the algebraically special mode and the neighboring branch cut, whereas the analogous feature is absent in the Regge-Wheeler response \cite{Besson:2026aog}. This provides a complementary illustration of why amplitudes, phases, and source parity contain information not encoded in the QNM frequencies alone.}
Our results suggest that even in events where the polar sector is dominant, the coupling can induce a considerable amplitude in the axial sector. For example, in the cases shown in Fig.~\ref{fig:amplitude}, the amplitude of the axial sector is $\sim 10\%$ of the amplitude of the even sector.

\section{Conclusions}\label{sec:conclusion}

We investigated QNM excitation and mixing in coupled systems within the parametrized QNM framework~\cite{Cardoso:2019mqo,McManus:2019ulj,Kimura:2020mrh,Volkel:2022aca,Cardoso:2024mrw}. Our analysis goes beyond the usual characterization of deviations from GR in terms of shifts of the QNM frequencies, because it quantifies how the coupling to additional degrees of freedom modifies the mode content and, in particular, the excitation amplitudes appearing in the ringdown signal. To this end, we combined frequency-domain calculations based on DI and a matrix-valued CF method with numerical time-domain evolutions, from which we extract the QNM frequencies and amplitudes using Prony and least-squares methods.

In the weak-coupling regime, the structure of the coupled equations provides a simple perturbative interpretation of mode excitation. Starting from initial data entirely contained in the primary sector, the secondary field is sourced at first order in the coupling, whereas its backreaction onto the primary field enters at second order. Accordingly, the amplitudes measured in the secondary-field waveform scale linearly with the coupling parameter $\alpha$ at leading order, while the coupling-induced corrections and the secondary-led contribution to the primary waveform scale as $\alpha^2$. A similar scaling was explored in  Ref.~\cite{Asuncion:2026efk}, where the authors studied gravitational-scalar perturbations in shift-symmetric Horndeski gravity. Our time-domain calculations extended this hierarchical scaling to both scalar and vector secondary fields over a broad range of parametrized couplings characterized by different radial profiles, i.e., different values of the index $j$. We provided numerical coefficients for these amplitude expansions up to $j=15$ that can be used to estimate the order of magnitude of mode mixing for couplings with different radial profiles.

An important consequence of these results is that the phenomenology of coupled systems cannot, in general, be characterized by frequency shifts of the primary modes alone. Unlike single-field modifications, the coupling induces mode mixing that excites an additional family of secondary-led QNMs. In the parameter space explored here, while the frequency shifts of the fundamental mode remain moderate, this additional family can acquire a non-negligible excitation amplitude in the gravitational sector; see also 
Ref.~\cite{Lestingi:2025jyb}.
The excitation amplitudes may thus constitute a complementary, and in some situations potentially more sensitive, probe of additional degrees of freedom. This is particularly relevant for black hole spectroscopy, where such modes could compete with other subdominant contributions to the ringdown, such as linear overtones and nonlinear or quadratic QNMs. Hence, a consistent interpretation of high-SNR ringdown signals may require modeling not only changes in the QNM spectrum, but also changes in the relative excitation of different families of modes.

As a theory-specific application of our calculations, we mapped our parametrized description to dCS gravity. In this theory, the coupling between axial gravitational perturbations and the scalar field directly corresponds to one of the parametrized coupling functions considered here. Consequently, we can translate the numerical amplitude coefficients from our time-domain evolutions into concrete estimates of the excitation amplitudes in dCS gravity. This example illustrates how our theory-agnostic framework connects generic ringdown phenomenology to specific modified gravity theories.

We also investigated the qualitatively different situation in which the coupled fields are nearly degenerate, using the Zerilli/Regge-Wheeler system as an example. Since these sectors are isospectral in the GR limit, 
the standard small-coupling hierarchy is no longer sufficient to describe the excitation pattern. Even for small couplings, the secondary sector can be appreciably excited, and the coupling breaks 
the degeneracy of the two QNM families. In this situation, as the two QNM families are nearly degenerate in frequency, the signal separation depends on the relative amplitudes, phases, and parity content. This shows the importance of mode excitation as an observable in situations involving nearly resonant QNM spectra.

Beyond the QNM-dominated regime, our time evolutions also revealed an interesting dependence of the late-time response on the radial structure of the coupling. Couplings with larger values of $j$, which are increasingly localized toward the near-horizon region, can significantly modify the onset and amplitude of the late-time tail and extend the interval over which the waveform is dominated by damped oscillations.
This behavior suggests an interplay between the location of the coupling relative to the effective-potential barrier and the propagation of the low-frequency components responsible for the late-time response. A more systematic study of this effect and of its analytical interpretation is left for future work.

An obvious extension of our work would be to consider the observationally more relevant, but harder to model, case of perturbations of rotating black holes beyond GR.
The parametrized QNM framework has been used to describe deviations from the Teukolsky equation~\cite{Li:2022pcy,Hussain:2022ins,Cano:2024jkd,DeSimone:2026xnz}, and it was recently extended to the time domain~\cite{DeSimone:2026waz}.
Even in the single-field case, rotation can induce coupling between different angular harmonics, making the type of coupled analysis developed in this work particularly relevant for practical applications.
A natural direction for future research would therefore be to incorporate secondary fields and investigate the effects of field coupling in rotating black hole spacetimes.
Ultimately, the resulting parametrized coupling coefficients could be constrained using gravitational-wave observations, extending parametrized ringdown studies of current data~\cite{Carullo:2021dui,Pompili:2025cdc,Grimaldi:2026prn,Volkel:2026qqz} 
as well as predictions for next-generation detectors~\cite{Maselli:2023khq,Toubiana:2023cwr}.

\acknowledgments

C.F.B.M. and O.R.M. acknowledge Coordenação de Aperfeiçoamento de Pessoal de Nível Superior–Brasil (CAPES)–Finance Code 001, Fundação Amazônia de Amparo a Estudos e Pesquisa (FAPESPA), and Conselho Nacional de Desenvolvimento Científico e Tecnológico (CNPq).
E.B. is supported by NSF Grants No.~AST-2513337, No.~AST-2307146, No.~PHY-2513337, No.~PHY-090003, and No.~PHY-20043, by John Templeton Foundation Grant No.~62840, by the Simons Foundation [MPS-SIP-00001698, E.B.], by the Simons Foundation International [SFI-MPS-BH-00012593-02], and by Italian Ministry of Foreign Affairs and International Cooperation Grant No.~PGR01167.
This work was carried out at the Advanced Research Computing at Hopkins (ARCH) core facility (\url{https://www.arch.jhu.edu/}), which is supported by the NSF Grant No.~OAC-1920103.
S.\,H.\,V. acknowledges funding from the European Union’s Horizon MSCA-2022 research and innovation programme “EinsteinWaves” under grant agreement No.~101131233.

\appendix

\section{Consistency of the frequency-domain calculations}\label{app:frequencies}

\begin{figure} [t]
    \centering
    \includegraphics[width=1\linewidth]{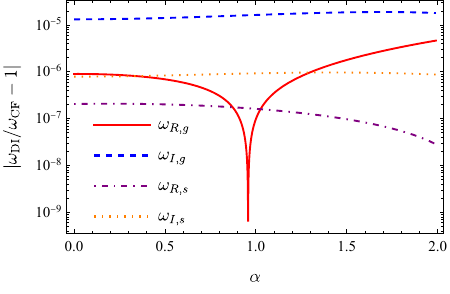}
    \caption{Comparison between the DI and CF methods as a function of the coupling parameter $\alpha$, for the case $j=3$. The two methods yield consistent results across the region of parameter space we considered. 
    }
    \label{fig:mode_com}
\end{figure}

In Fig.~\ref{fig:mode_com}, we show the relative difference between fundamental QNMs computed using DI and the CF method for the coupled system of gravitational and scalar perturbations. As expected, the coupling gives rise to two families of modes, usually referred to as the primary-led ($\omega_{s_1}=\omega_{r,s_1}+i\omega_{i,s_1}$) and secondary-led ($\omega_{s_2}=\omega_{r,s_2}+i\omega_{i,s_2}$) branches.
We show the case $j=3$, although we found a similar behavior other values of $j$ and for different spins. As can be seen, the fundamental modes obtained with the two methods are in good agreement.

\section{Benchmarking the Prony method}\label{app:Benchmarking}

\begin{figure*}
    \centering
    \includegraphics[width=0.99\linewidth]{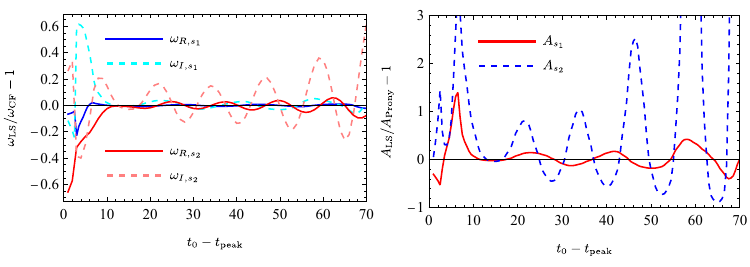}\\
    \includegraphics[width=0.99\linewidth]{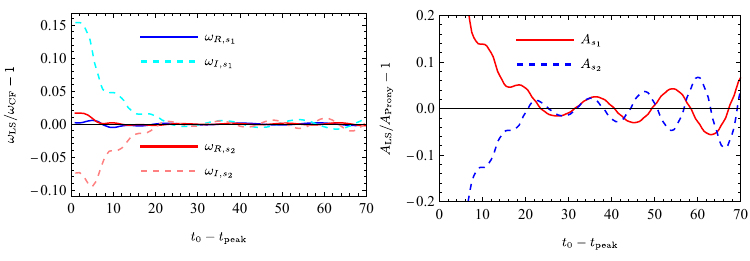}
    \caption{Comparison between the QNM frequencies computed with the CF method and those extracted from time-domain evolutions via a least-squares fit, as a function of the fitting start time relative to the peak, for $\alpha=2$ and $j=3$. The upper panels refer to the primary signal, while the lower panels refer to the secondary signal. The left panels compare the oscillation frequencies obtained with the two methods, whereas the right panels show the deviation of the amplitudes extracted with the least-squares fit relative to the corresponding Prony values. For the gravitational signal, we find that $t_0-t_{\rm peak}=15M$ provides an appropriate choice for the beginning of the ringdown window, as it yields good agreement in both frequency and amplitude. For the scalar signal, the frequency is also well captured near this value, although the amplitude approaches the Prony result more closely for slightly later starting times.}
    \label{fig:frequency_time_comparison}
\end{figure*}

In this Appendix, we benchmark the Prony method against the least-square technique. We focus on the case of a gravito-scalar coupling with $\alpha=2$ and $j=3$. 
In our analysis, the Prony method yields frequency estimates that are consistently in good agreement with the CF results, while exhibiting greater numerical stability than the least-squares technique over a broader range of ringdown windows. This is because the Prony method effectively linearizes the extraction of frequencies via singular value decomposition, avoiding the local minima and parameter degeneracies that often affect nonlinear least-squares fits in the presence of multiple damped modes. However, we also use the least-squares method to extract the frequencies, amplitudes, and phases present in the signal as an independent check. 
For the least-squares technique, the left panels of Fig.~\ref{fig:frequency_time_comparison} show the frequencies extracted from the time-domain signals, compared with those obtained through the CF method described in Sec.~\ref{sec:frequencydomain}. We find that for the primary signal (top-left panel) taking the initial time to be $15M$ after the peak yields the best agreement between the time-domain and frequency-domain results. This provides an initial indication of a suitable starting point for the ringdown window in the primary signal. For the secondary signal, the starting value is around $22M$.

We also analyze the amplitude at the peak as extracted with the least-squares method. Interestingly, the Prony method yields an approximately constant amplitude over a wide range of initial times, which we use as a reference to benchmark the least-squares results. In the right panels of Fig.~\ref{fig:frequency_time_comparison}, we show the deviation from the Prony value. Once again, we find that the amplitudes extracted with the least-squares method are stable for the gravitational signal when the fitting window begins around $15M$, while for the scalar signal they approach the Prony result more closely around $22M$. In both cases, the two methods are consistent, with the amplitudes extracted from the least-squares fit oscillating around the ones obtained from the Prony method.

\begin{figure*}
    \includegraphics[width=\linewidth]{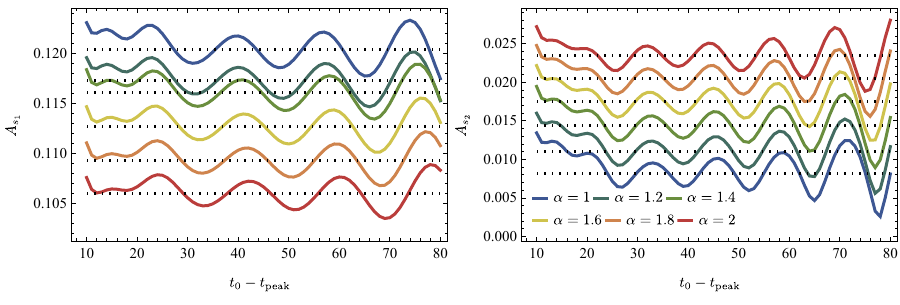}\\
    \includegraphics[width=\linewidth]{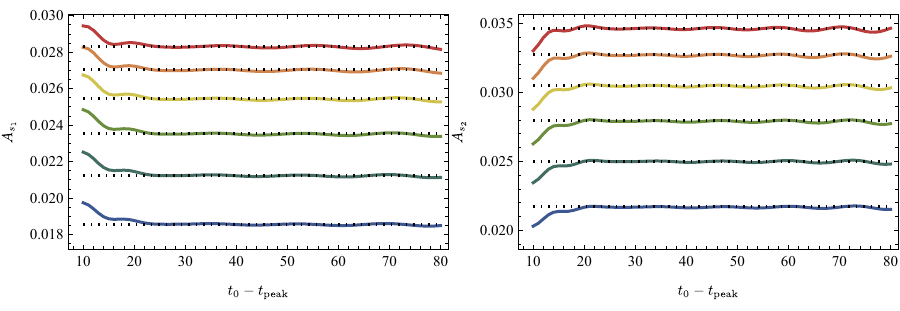}
    \caption{Comparison between the amplitudes extracted with a least-squares fit (solid curves), using the QNM frequencies computed with the CF method in the frequency domain, and those obtained with the Prony method (dotted horizontal lines). We consider here the coupled primary-secondary system with $j=3$, with $\alpha$ in the range $[1,\,2]$. The upper panels show the amplitudes extracted from the gravitational signal, while the lower panels show those extracted from the scalar signal. The left panels correspond to the gravitational-led mode, whereas the right panels correspond to the scalar-led mode.}
    \label{fig:amps_alpha_lq}
\end{figure*}

To further validate the Prony method against the least-squares fit, we also compute the amplitude using the least-squares method while fixing the fundamental frequencies of the gravitational and scalar fields. We perform this analysis for several values of $\alpha$ in the interval $[1,\,2]$ (shown by the solid curves ranging from blue to red) and for different choices of the initial time $t_0$. The results are displayed in Fig.~\ref{fig:amps_alpha_lq}. In all cases, the least-squares method yields amplitudes that oscillate around the values obtained with the Prony method, represented by the dashed lines. Interestingly, the scalar field perturbation, which is induced by the coupling between the two fields, exhibits a cleaner behavior in the least-squares extraction.

It is interesting to look at the dependency of the extracted parameters using the single damped sinusoid ansatz as a function of starting time shown in Figs.~\ref{fig:frequency_time_comparison} and~\ref{fig:amps_alpha_lq}.
The oscillatory patterns are not a numerical artifact or problem of the fitting procedure; instead, they are expected from linear-signal analysis.
In general, the analyzed time-domain signal is not completely described by the damped-sinusoid model, which inherently introduces biases in the extracted parameters (i.e., systematic errors).
We have verified that across different fitting windows the reference values obtained using the Prony method vary substantially less than those obtained from least-squares fits, which track only the fundamental modes.

Qualitatively similar oscillatory patterns have recently been observed in Ref.~\cite{Volkel:2025jdx}, which are reminiscent of the best-fit frequencies found in this work.
While oscillatory patterns at early times may point towards overtones and the prompt response, the late-time behavior should in principle be well approximated by the fundamental mode.
However, in our system, there are two possible types of late-time contributions.
First, the long-range nature of the perturbation potentials gives rise to power-law tails, most prominently the Price tail. The tail is not dominant at intermediate times, but it may introduce visible systematic errors (see Ref.~\cite{Volkel:2025jdx}).
Second, when two or more fields are coupled, the QNMs associated with each field, whether scalar-, vector-, axial-, or polar-led, are expected to leave an imprint on the oscillations of the others.
Consequently, the fundamental modes of the coupled fields should be present in the analyzed signal, even if only with small amplitudes.
Since the damping times of the primary and secondary fundamental mode are similar, one should expect that if one of the fields contaminates the other at some time, it remains relevant throughout the ringdown.
Thus, the presence of oscillatory patterns as a function of the starting time is a clear trace of potentially unknown features.

\clearpage

\bibliography{refs}
\end{document}